\documentstyle[12pt]{article}
\begin{document}
\title{A new massive vector field theory}
\author {{\small Zhiyong Wang$^1$ and Ailin
Zhang$^2$}\\
{\small $^1$ P. O. Box 1, Xindu, Chengdu, Sichuan, 610500, P. R. China}\\ 
{\small $^2$ Institute of theory Physics, P. O. Box 2735, Beijing, 100080, 
P. R. China}\\}
\date {}
\maketitle
\begin{center}
\begin{abstract}
In this paper, we put forth a new massive spin-1 field theory. In contrast
to the quantization of traditional vector field, the quantization of the
new vector field is carried out in a natural way. The Lorentz
invariance of the theory is discussed, where owing to an interesting
feature of the new vector field, the Lorentz invariance has a special
meaning. In term of formalism analogical to QED(i. e. spinor QED), we
develop the quantum electrodynamics concerning the new spin-1 particles,
say, vector QED, where the Feynman rules are given. The
renormalizability of vector QED is manifest without the aid of Higgs
mechanism. As an example, the polarization cross section $\sigma_{polar}$
for $e^+ e^-\to f^+ f^-$ is calculated in the lowest order. It turns
out that $\sigma_{polar}\sim 0$ and the momentum of $f^+$ and $f^-$ is
purely longitudinal.
\end{abstract}
\end{center}
\section{Introduction}
\indent
\par After the introduction of Dirac's equation\cite{dirac}, the
search began for similar equations for higher spins. In the past, various
approaches have been tried$\to$ equations describing many masses and spins
particle, non-Lagrangian theories and theories with indefinite
metric\cite{bha} {\it et al}. At first, it was observed that, apart from
spin-1/2, none of the other spins obeys a single-particle relativistic
wave equation. For example, it was generally believed that for spins 0 and
1, the Klein-Gordon\cite{klein}-\cite{gordon} and Proca 
equations\cite{proca} were unique, respectively. However, more than 60
years ago, it was found that the
Kemmer-Duffin-Petian\cite{kemmer}-\cite{petiau} equations(KDF equations)
can describe both spin-0 and spin-1 objects. Since then, many
more systems of equations for arbitrary spins, which originate from
different assumptions after considering their invariance under Lorentz
group, have been found.
\par Unfortunately, it has been known for a long time that there still
exist many difficulties in the construction of higher spins field
theories, which has turned out to be
the most intriguing and challenging in theoretical physics. Especially,
such a theory has touched upon some of the most basic
ingredients of present-day physical theory. For example, in those
theories, either the usual connection between spin and statics is
violated, or the law of causation do not hold, or the negative energy
difficulty is still encountered after second quantization having been
accomplished, or, in the presence of interactions, the complex
energy eigenvalues, superluminal propagation of waves and many other
undesirable features\cite{velo}-\cite{pereira} were found too. In
particular, these behaviors were exhibited by both the KDP and the Proca
equations\cite{seet}.
\par On the other hand, both the Maxwell equations for electromagnetic
field and Yang-Mills equations for non-Abelian gauge field\cite{ym} can be
restated in terms of a spinor notation\cite{opp}-\cite{mori} resembling
the one for Dirac equation, which motivates people to extend the idea
to a massive system with arbitrary spin\cite{caian}-\cite{hurl}. In this
paper, we set up a new massive vector field equation(called as Dirac-like 
equation), which takes a form similar to the Dirac equation but involves
some six-by-six matrices. The equation is no longer equivalent to any 
other existed ones such as the KDP equation(in the spin-1 case), the Proca
equation and the Weinberg equation\cite{bha}, \cite{joos}, etc. Moreover,
it is observed, on one hand, the general solution of a
relativistic wave equation can be, not only the sum of positive-frequency
and negative-frequency parts(denoted by $\varphi_1$), but also the
difference of them(denoted by $\varphi_2$). On the other hand, the
positive-frequency and negative-frequency solutions of
an equation are linearly independent such that $\varphi_1$ and $\varphi_2$
transform in the same way under the Lorentz and the gauge transformations,
Therefore, these two types of general
solutions($\varphi_1$ and $\varphi_2$) are simultaneously used to
construct the Lagrangian of the vector field. As a result, all those
difficulties mentioned above are swept away.
\par The paper is organized as follows. In Sec. II, the Dirac-like
equation is put forward and the corresponding plane wave solution is
discussed. In Sec. III, by choosing a suitable Lagrangian, from which the
Dirac-like equation can be derived, we quantilize the Dirac-like field
naturally according to Bose-Einstein statistics, where the energy is
positive-definite too. In Sec. IV, an interesting feature of the field is
shown. From the Lagrangian, we construct the Feynman propagator for
the field in Sec. V, with the causality being preserved. In Sec. VI, the
Lorentz invariance of the theory is discussed, where the action of the
transversal field is proved to be Lorentz invariant. In Sec. VII, we
develop the Feynman rules for the vector QED, where, as an example, the
polarization cross section for the process $e^+ e^-\to f^+ f^-$ is
calculated. At last, in Sec. VIII, we show that vector QED is a
renormalizable theory. The system of natural units and Bjorken
conventions\cite{bjor} are used throughout in the paper.
\section{A new massive vector field equation} 
\indent 
\par In analogy with the construction of Dirac equation, we
can set up the following free relativistic "Dirac-like" equation as
follow:
\begin{eqnarray}\label{dl}
(i\beta^\mu \partial_\mu-m)\varphi(x)=0,
\end{eqnarray}
where $m$ is the mass and  
$\beta^\mu=(\beta^0,\vec{\beta})$ satisfies
\begin{eqnarray}\label{be}
\left\{\begin{array}{ll}
&(\vec{\beta}\cdot \vec{p})^3=-\vec{p}^2(\vec{\beta}\cdot \vec{p}) \\
&\beta^0\beta^i+\beta^i\beta^0=0 \\
&(\beta^0)^2=1 ,
\end{array}
\right.
\end{eqnarray}
where $\vec{p}$ is an arbitrary three dimensional vector such as the
spatial component of a 4-momentum. To express the $\beta$ matrix
explicitly, we choose 
\begin{eqnarray}\label{ma}
\begin{array}{cc}
\beta^0=\left (
\begin{array}{lcr}
I_{3\times 3} & 0\\
0 & -I_{3\times 3}
\end{array} \right ), &
\vec{\beta}=\left (
\begin{array}{lcr}
0 & \vec{\tau} \\
-\vec{\tau} & 0
\end{array} \right ),
\end{array}
\end{eqnarray}
where $I_{3\times 3}$ is the $3\times 3$ unit matrix 
and $\vec{\tau}=(\tau_1, \tau_2, \tau_3)$, in which
\begin{eqnarray}
\begin{array}{ccc}
\tau_1=\left (
\begin{array}{lcr}
0 & 0 & 0 \\
0 & 0 & -i \\
0 & i & 0
\end{array} \right ), &
\tau_2=\left (
\begin{array}{lcr}
0 & 0 & i \\
0 & 0 & 0 \\  
-i & 0 & 0
\end{array} \right ), &
\tau_3=\left (
\begin{array}{lcr}
0 & -i & 0 \\
i & 0 & 0 \\
0 & 0 & 0
\end{array} \right ).
\end{array}
\end{eqnarray}
It is easy to check that  $\beta^\mu$ given above
satisfy the relation (\ref{be}). Let $\vec{\alpha}=\beta^0\vec{\beta}$,
Eq. (\ref{dl}) can be rewritten as
\begin{eqnarray}\label{dl2}
i\partial_t\varphi (x)=(\vec{\alpha}\cdot \hat{p}+\beta^0m)\varphi(x),
\end{eqnarray} 
where $\hat{p}=-i\nabla$ is the momentum operator. Accordingly, the
Hamiltonian is $\hat{H}=\vec{\alpha}\cdot \hat{p}+\beta^0m$. Let 
$\hat{L}=\vec{x}\times \hat{p}$ represent the operator of orbit angular
momentum, we have
\begin{eqnarray}
\begin{array}{cc}
[\hat{H},\hat{L}]=-i\vec{\alpha}\times \hat{p}, &
[\hat{H},\hat{L}+\vec{S}]=0,
\end{array}
\end{eqnarray}
where
\begin{eqnarray}
\vec{S}=\left (
\begin{array}{lcr}
\vec{\tau} & 0 \\
0 & \vec{\tau}
\end{array} \right )
\end{eqnarray}
is the spin matrix. Since $\vec{S}^2=2$, the corresponding
particle(called Dirac-like particle) has spin 1, which will be
demonstrated further in Sec. VI. Namely, Eq. (\ref{dl}) represents the
equation of a massive vector field. As we will see later, the
Dirac-like field is different from any other vector fields, and
hence is a completely new one.
\par Substituting the plane wave solution $\varphi (p)e^{-ipx}$ into Eq.
(\ref{dl2}), we obtain the representation of Eq. (\ref{dl2}) in momentum 
space as follow:
\begin{eqnarray}\label{dl3}
(E-\vec{\alpha}\cdot \hat{p}-\beta^0m)\varphi (p)=0,
\end{eqnarray}  
where $E=p^0$ is the energy. When $\{\hat{H}, \hat{p},
{\vec{s}\cdot\hat{p} \over |\vec{p}|} \}$ are chosen 
as dynamical completeness operators, the fundamental solutions of Eq. 
(\ref{dl3}) are derived
\begin{eqnarray}\label{f1}
\begin{array}{cc}
|\vec{p},s\rangle_+= \sqrt {{E_s+m\over 2m}} \left (
\begin{array}{cr}
\eta_s \\
{\vec{\tau}\cdot\vec{p} \over E_s+m}\eta_s
\end{array} \right ), &
|\vec{p},s\rangle_-= \sqrt {{E_s+m\over 2m}} \left (
\begin{array}{cr}
{-\vec{\tau}\cdot\vec{p} \over E_s+m}\eta_s \\
\eta_s
\end{array} \right ),
\end{array}
\end{eqnarray}
where $|\vec{p},s\rangle_+$ and $|\vec{p},s\rangle_-$ correspond to the
positive and the negative energy solutions, respectively. $s=-1, 0, 1$ and
\begin{eqnarray}\label{e}
E_{-1}=E_{1}=E_\perp=\sqrt{\vec{p}^2+m^2}, & E_0=E_L=m,
\end{eqnarray}   

\begin{eqnarray}\label{et}
\eta_{1}={1 \over \sqrt{2}|\vec{p}|} \left (
\begin{array}{lcr}
{p_1p_3-ip_2|\vec{p}| \over p_1-ip_2} \\
{p_2p_3+ip_1|\vec{p}| \over p_1-ip_2} \\
-(p_1+ip_2)
\end{array} \right ), &
\eta_{-1}=\eta_{1}^\ast , &
\eta_0={1\over |\vec{p}|} \left (
\begin{array}{lcr} 
p_1 \\
p_2 \\
p_3
\end{array} \right ).
\end{eqnarray}
where $\eta_{1}^\ast$ is the complex conjugate of $\eta_1$. Let
$\eta^\dagger_s$ stands for the Hermitian conjugate of $\eta_s$, we have
\begin{eqnarray}\label{orth}
\left\{\begin{array}{ll}
&\eta^\dagger_s\eta_{s'}=\delta_{ss'} \\
&\sum\limits_{s}\eta_s\eta^\dagger_s=I_{3\times 3} ,
\end{array}
\right.
\end{eqnarray} 
which means that $\{\eta_s\}$ form a complete orthonormal basis. It
can be verified that
\begin{eqnarray}
{1\over |\vec{p}|}\vec{\tau}\cdot \vec{p}\eta_s=\lambda_s \eta_s,
\end{eqnarray} 
where $\lambda_1=1$, $\lambda_{-1}=-1$ and $\lambda_0=0$. Obviously,
the solutions with $s=\pm 1$ correspond to the transversal polarization
solutions of Eq. (\ref{dl3}),  while another one corresponds to the
longitudinal polarization solution. Besides, an interesting result can be
read from Eq. (\ref{e}): the transversal polarization particles have
energy $\sqrt{\vec{p}^2+m^2}$ while the longitudinal one has energy m,
whose physical reason will be explained in Sec. IV.
\par As for the wave functions, the ones in momentum space are chosen as
\begin{eqnarray}
\chi(\vec{p},s)=|\vec{p},s\rangle_+, &
y(\vec{p},s)=|-\vec{p},s\rangle_-=\sqrt{E_s+m \over 2m} \left (
\begin{array}{cr}
{\vec{\tau}\cdot \vec{p}\over E_s+m}\eta_s \\
\eta_s
\end{array} \right ).
\end{eqnarray}
and the corresponding ones in position space are 
\begin{eqnarray}
\varphi_{p,s}(x)=\sqrt{{m\over vE_s}} \chi (\vec{p},s)e^{-ipx}, 
~~\varphi_{-p,s}(x)=\sqrt{{m\over vE_s}} y (\vec{p},s)e^{ipx}.
\end{eqnarray}
Taking into account Eq. (\ref{orth}), we have(here
${\bar\chi}=\chi^\dagger\beta^0$, and so on)
\begin{eqnarray}\label{orth1}
\left\{\begin{array}{ll}
&\bar\chi(\vec{p},s)\chi(\vec{p},s')=-\bar
y(\vec{p},s)y(\vec{p},s')=\delta_{ss'} \\
&\bar\chi(\vec{p},s)y(\vec{p},s')=\bar 
y(\vec{p},s)\chi(\vec{p},s')=0 .
\end{array}
\right.
\end{eqnarray}
\begin{eqnarray}\label{sum1}
\left\{\begin{array}{ll}
&\sum\limits_{s}\chi(\vec{p},s)\bar\chi(\vec{p},s)={\beta\cdot p_{\bot}+m
\over 2m}(I_{2\times 2}\bigotimes\sum\limits_{s=\pm 1}\eta_s\eta^\dagger_s)
+{\beta\cdot p_L+m \over 2m}(I_{2\times 2}\bigotimes\eta_0\eta^\dagger_0)
\\
&\sum\limits_{s}y(\vec{p},s)\bar y(\vec{p},s)={\beta\cdot p_{\bot}-m
\over 2m}(I_{2\times 2}\bigotimes \sum\limits_{s=\pm 1}\eta_s\eta^\dagger_s)
+{\beta\cdot p_L-m \over 2m}(I_{2\times 2}\bigotimes
\eta_0\eta^\dagger_0).
\end{array}
\right.
\end{eqnarray}
where $\bigotimes$ is the direct product symbol. Obviously, Eq.
(\ref{orth1}) and Eq. (\ref{sum1}) complete the orthonormality relations
and the spin summation relations for the massive vector field, from which
the completeness relation can be derived too.
\section{Quantization of the Dirac-like field}
\indent
\par In this section, we will quantize the Dirac-like field in a manner
similar to that used for Dirac field but obeying the Bose-Einstein
statistics. Two types of general solutions
of Eq. (\ref{dl}) can be constructed simultaneously as follows:
\begin{eqnarray}\label{vr1}
\varphi_1(x)\equiv \varphi_+(x)+\varphi_-(x)={\sum\limits_{\vec{p},s} 
[a(\vec{p},s)\varphi_{p,s}(x)+b^\dagger (\vec{p},s)\varphi_{-p,s}(x)
}],\\
\label{vr2}
\varphi_2(x)\equiv \varphi_+(x)-\varphi_-(x)={\sum\limits_{\vec{p},s}
[a(\vec{p},s)\varphi_{p,s}(x)-b^\dagger (\vec{p},s)\varphi_{-p,s}(x) }] ,
\end{eqnarray}
where 
\begin{eqnarray}
\varphi_+(x)={\sum\limits_{\vec{p},s}a(\vec{p},s)\varphi_{p,s}(x)}, &
\varphi_-(x)={\sum\limits_{\vec{p},s}b^\dagger (\vec{p},s)\varphi_{-p,s}(x)},
\end{eqnarray}
are the positive and the negative frequency parts of the general
solutions, respectively, and $a(\vec{p},s)$, $b^\dagger (\vec{p},s)$ are 
coefficients.
\par To deduce the free Dirac-like equation, the free Lagrangian is
chosen as
\begin{eqnarray}\label{lag1}
L={\bar \varphi}_2(x)(i\beta^\mu\partial_\mu-m)\varphi_1(x).
\end{eqnarray}
Then the canonical momentum conjugates to $\varphi_1(x)$ is
\begin{eqnarray}
\pi(x)=\partial L/\partial \dot{\varphi_1}=i\varphi^\dagger_2(x),
\end{eqnarray}
and thus, the Hamiltonian $H$ and the momentum $\vec{p}$ are,
respectively, 
\begin{eqnarray}\label{hp}
\begin{array}{c}
H=\int{[\pi(x)\dot{\varphi}_1(x)-L]d^3x}=\int{\varphi^\dagger
_2(-i\vec{\alpha}\cdot
\nabla+\beta^0m)\varphi_1(x)d^3x}, \\
\vec{p}=-\int{\pi(x)\nabla\varphi_1
d^3x}=-i\int{\varphi^\dagger _2(x)\nabla\varphi_1 d^3x}.
\end{array}
\end{eqnarray}
\par Now, we promote $\varphi_1(x)$ and $\pi(x)$ to operators and
the canonical equal time commutation relations become
\begin{eqnarray}
\begin{array}{ll}
[\varphi_{1\alpha}(\vec{x},t),\pi_\beta(\vec{x}',t)] =
i\delta_{\alpha\beta}\delta^3(\vec{x}-\vec{x}'), 
\end{array}
\end{eqnarray}
with the others vanishing. In term of $a(\vec{p},s)$ and $b(\vec{p},s)$,
we get the following commutation relations 
\begin{eqnarray}\label{comm}
[a(\vec{p},s), a^\dagger (\vec{p}',s')]=[b(\vec{p},s),
b^\dagger (\vec{p}',s')]=\delta_{\vec{p}\vec{p}'}\delta_{ss'},
\end{eqnarray}
and all other commutators vanish. Making use of Eq. (\ref{orth1}),
(\ref{vr1}), (\ref{vr2}) and (\ref{comm}), Eq. (\ref{hp}) transforms into 
\begin{eqnarray}
\begin{array}{l}
H={\sum\limits_{\vec{p},s}}E_s[a^\dagger(\vec{p},s)a(\vec{p},s)+
b^\dagger(\vec{p},s)b(\vec{p},s)+1], \\
\vec{p}={\sum\limits_{\vec{p},s}}[a^\dagger(\vec{p},s)a(\vec{p},s)+
b^\dagger(\vec{p},s)b(\vec{p},s)],
\end{array}   
\end{eqnarray}
where $a^\dagger(\vec{p},s)$, $a(\vec{p},s)$ are the creation and
annihilation operators of particles, respectively, while $b^\dagger
(\vec{p},s)$, $b(\vec{p},s)$ are the corresponding ones of antiparticles.
Obviously, the Dirac-like field obeys the
Bose-Einstein statics and the corresponding energy is
positive-definite, which is just what we desire.
\section{Character of the new vector field}
\indent
\par Let us turn our attention to the expectation value $\vec{v}$
of the velocity operator $\dot{\vec{x}}=i[\hat{H},\vec{x}]=\vec{\alpha}$
\begin{eqnarray}\label{exp}
\vec{v}=\int{\varphi^\dagger _2(x)\dot{\vec{x}}\varphi_1(x)d^3x}=
\int{\varphi^\dagger _2\vec{\alpha}\varphi_1d^3x}.
\end{eqnarray}
In fact, $\varphi^\dagger _2\vec{\alpha}\varphi_1$ can be regarded as the 
probability current density and hence $\vec{v}$ corresponds to the
probability current. Substituting Eq. (\ref{var1}), (\ref{var2}) into Eq.
(\ref{exp}), we obtain
\begin{eqnarray}
\vec{v}=\vec{v}_\bot + \vec{v}_L,
\end{eqnarray}
where $\vec{v}_L$ represents the current related to transversal particle
only, while $\vec{v}_{\bot}$ corresponds to the current with the
contribution of longitudinal particles involved in too. The explicit
expression of them are 
\begin{eqnarray}\label{cur}
\begin{array}{ll}
\vec{v}_L &=\sum\limits_{\vec{p}}\sum\limits_{s=\pm 1}{\vec{p}\over E}
[a^\dagger (\vec{p},s)a(\vec{p},s)-b^\dagger (\vec{p},s)b(\vec{p},s)] \\
&+\sum\limits_{\vec{p}}\sum\limits_{s=\pm 1}(-1)^{{(s-1)\over 2}}
{m\over E}\vec{\eta}_0
[a^\dagger (\vec{p},s)b^\dagger
(-\vec{p},s)e^{i2Et}-a(\vec{p},s)b(-\vec{p},s)e^{-i2Et}],\\
\vec{v}_\bot &= 
\sum\limits_{\vec{p}}\sum\limits_{s=\pm 1}\sqrt{{1\over 2E(E+m)}}|\vec{p}|
\{[-a^\dagger (\vec{p},0)a(\vec{p},s)e^{i(m-E)t}
+b^\dagger (\vec{p},s)b(\vec{p},0)e^{-i(m-E)t}]\vec{\eta}_s + h.c.\} \\
&+\sum\limits_{\vec{p}}\sum\limits_{s=\pm 1}\sqrt{{E+m\over 2E}}
\{[a^\dagger (\vec{p},0)b^\dagger (-\vec{p},s)\vec{\eta}_s
+a^\dagger (\vec{p},s)b^\dagger (-\vec{p},0)\vec{\eta}^*_s]e^{i(m+E)t} -
h.c.\},
\end{array}
\end{eqnarray}
respectively, in which $E=E_\bot=\sqrt{\vec{p}^2+m^2}$ and $\vec{\eta}_s$
is the vector representation of Eq. (\ref{et}).
\par Before going on, we will give some discussions about Eq. (\ref{cur}):
\par (1) As far as $\vec{v}_L$ is concerned, the first term corresponds to
the classic current with group velocity ${\vec{p} \over E}$ of the
wave packet, while the second term corresponds to the
zitterbewegung\cite{schro} current of the transversal particle. The
zitterbewegung current does not vanish as $\vec{p}\to 0$, which infers
that it is intrinsic and independent of macroscopic classic motion.
\par (2) As for $\vec{v}_\bot$, the first term corresponds to the
current resulting from the interference between the transversal and the
longitudinal particle, and the second term corresponds to the
zitterbewegung current containing the contribution of longitudinal
particle.
\par (3) $\vec{\eta_0}$ is parallel to $\vec{p}$ while
$\vec{\eta}_{\pm 1}$ is vertical to $\vec{p}$(denoted by
$\vec{\eta_0}\parallel \vec{p}$ and $\vec{\eta_0}\perp \vec{p}$,
respectively), thus $\vec{v}_L\parallel \vec{p}$ while
$\vec{v}_\bot\perp \vec{p}$(or, $\vec{v}_\perp\cdot \vec{p}=0$), On the
other hand, the longitudinal particle does not contribute to
$\vec{v}_L$(which contributes only to $\vec{v}_\bot$). Therefore, the
longitudinal particle makes no contribution to the current in the
direction of momentum $\vec{p}$, which is consistent with the statement
that $E_L=m$ obtained earlier(see Eq. (\ref{e})). As a consequence, we can
regard the longitudinal particle as the one corresponding to standing
wave. In fact, the behaviors of the new vector field are similar to those
of the electromagnetic wave propagating in hollow metallic
waveguide\cite{jack}, where the longitudinal component of the
electromagnetic wave makes no contribution to the flow of energy(or the
Poynting vector) along the waveguide.
\section{The free propagator of the Dirac-like field}
\indent
\par The free propagator of Dirac-like field is defined as 
\begin{eqnarray}
iR_f(x_1-x_2)\equiv \langle 0|T\varphi_1(x_1)\bar \varphi_2(x_2)|0\rangle,
\end{eqnarray}
where T is the time order symbol. Considering Eq. (\ref{sum1}), we obtain
\begin{eqnarray}
iR_f(x_1-x_2)=iR_{f\bot}(x_1-x_2)+iR_{fL}(x_1-x_2),
\end{eqnarray}
where $iR_{f\bot}(x_1-x_2)$ is the free propagator of transversal field
and $iR_{fL}(x_1-x_2)$ is the longitudinal one, which read 
\begin{eqnarray}
iR_{f\bot}(x_1-x_2)=\int {d^4p \over (2\pi)^4}{i\Omega_\bot \over
p^2_0-E^2_\bot+i\varepsilon} e^{-ip\cdot (x_1-x_2)} \\\nonumber
iR_{fL}(x_1-x_2)=\int {d^4p \over (2\pi)^4}{i\Omega_L \over 
p^2_0-E^2_L+i\varepsilon} e^{-ip\cdot (x_1-x_2)} ,
\end{eqnarray}
respectively, where $\varepsilon$ is a infinitesimal real quantity,
$p_\mu=(p_0,-\vec{p})$ and
\begin{eqnarray}
\Omega_\bot=(i\beta\cdot\partial+m)A_\bot \\\nonumber
\Omega_L=(i\beta\cdot\partial+m)A_L ,
\end{eqnarray}
in which $A_\bot=I_{2\times 2}\bigotimes \sum\limits_{s=\pm 1}\eta_s
\eta^\dagger_s$ and $A_L=I_{2\times 2}\bigotimes \eta_0 
\eta^\dagger_0$. By applying $p^2_0-E^2_\bot=p^2-m^2, (\beta\cdot
p)^2A_\bot=p^2A_\bot$ and $p^2_0-E^2_L=p^2_0-m^2, (\beta\cdot 
p)^2A_L=p^2A_L$, we have
\begin{eqnarray}
iR_{f\bot}(x_1-x_2)=\int {d^4p \over (2\pi)^4}{iA_\bot \over
\beta\cdot p-m+i\varepsilon} e^{-ip\cdot (x_1-x_2)} \\\nonumber
iR_{fL}(x_1-x_2)=\int {d^4p \over (2\pi)^4}{iA_L \over
\beta\cdot p-m+i\varepsilon} e^{-ip\cdot (x_1-x_2)} .
\end{eqnarray}
Due to $A_\bot +A_L=1$, 
\begin{eqnarray}
iR_{f}(x_1-x_2)=\int {d^4p \over (2\pi)^4}{i\over
\beta\cdot p-m+i\varepsilon} e^{-ip\cdot (x_1-x_2)},
\end{eqnarray}
which takes a form similar to the free propagator of Dirac field but
involves the matrices $\beta^\mu$ instead of Dirac matrices
$\gamma^\mu$
\par The representation of $iR_{f}(x_1-x_2)$ in momentum space is
\begin{eqnarray}
iR_{f}(p)=iR_{f\bot}(p)+iR_{fL}(p)={i \over \beta\cdot p-m+i\varepsilon},
\end{eqnarray}
where
\begin{eqnarray}\label{pro1}
iR_{f\bot}(p)={i \over \beta\cdot p-m+i\varepsilon}A_\bot,
iR_{fL}(p)={i \over \beta\cdot p-m+i\varepsilon}A_L.
\end{eqnarray}
It is easy to find that
\begin{eqnarray}
(i\beta\cdot\partial-m)R_{f}(x_1-x_2)=\delta^4(x_1-x_2).
\end{eqnarray}
Namely, $R_{f}(x_1-x_2)$ is the Green's function of free Dirac-like
equation. To make Eq. (\ref{pro1}) more explicit, we choose a frame in
which $\vec{p}=(0,0,p_3)$, then
\begin{eqnarray}\label{ma}
\begin{array}{cc}
A_\bot=I_{2\times 2}\bigotimes\left (
\begin{array}{lcr}
1 & 0 & 0\\
0 & 1 & 0\\
0 & 0 & 0
\end{array} \right ),
A_L=I_{2\times 2}\bigotimes\left (
\begin{array}{lcr}
0 & 0 & 0 \\
0 & 0 & 0 \\ 
0 & 0 & 1
\end{array} \right ),
\end{array}
\end{eqnarray}
\section{Lorentz invariance of the theory}
\indent
\par Considering the fact that the positive and the negative frequency
parts of Dirac-like field are linearly independent, one can readily
verify that $\varphi_1(x)$ and $\varphi_2(x)$(given by Eq. (\ref{vr1}) and 
(\ref{vr2}), respectively) transform in the same way under a Lorentz
transformation. 
\par In following special cases, Lorentz boost along the direction of the
Dirac-like field's motion or Lorentz boost $L(\vec{p})$ which makes a
Dirac-like particle from rest to momentum $\vec{p}$, the Lagrangian
$L_\varphi$ given by Eq. (\ref{lag1}) is easily proved to be Lorentz
invariant.
\par Now, we consider the variation of Lagrangian under an arbitrary
Lorentz transformation. Before embarking on the process, the
following fact should be noted: the longitudinal field makes no
contribution to the current in the direction of the momentum of Dirac-like
field and the energy of it is $E_L=m$, so we regard it as the one that
exists in a standing wave form or in a virtual form. Meanwhile,
for the independence of the longitudinal field with the interaction
related to
the current, the longitudinal field is taken as an unobservable one.
Certainly, the unobservable longitudinal field in one frame can be turned
into the observable transversal field in another frame and vice versa. 
However, the action of the transversal field is Lorentz invariant(just as
we will show later). As far as the noncovariance of the transversal
condition for the transversal field is concerned, it won't do any hurt to
the theory. The Lorentz invariant transition amplitude\cite{wein} still
can be obtained from a noncovariant Hamiltonian formulation of field
theory, just as what we have done in electromagnetic field theory with the
noncovariant Coulomb gauge. For the reasons mentioned above, we will
demonstrate only the invariance of the action of transversal field in
the following.
\par For generality, let us consider the case that the Lagrangian contains
an interaction term, in which the Dirac-like field is coupled to the
electromagnetic field in the minimal form, namely
\begin{eqnarray}\label{lag2}
L_\varphi={\bar \varphi_2}(x)[i\beta D_\mu-m]\varphi_1(x),
\end{eqnarray}
where $\varphi_1$ and $\varphi_2$ contain only the transversal parts, 
$D_\mu=\partial_\mu-ieA_\mu$, e is a dimensionless coupling
constant and $A_\mu$ is the electromagnetic field. The
Lorentz invariant free term for $A_\mu$ is omitted. Under Lorentz
transformation
\begin{eqnarray}
x^\mu\to x'^\mu=a^{\mu\nu}x_\nu, & D_\mu\to D'_\mu=a_{\mu\nu}D^\nu, &
d^4x\to d^4x'=d^4x,
\end{eqnarray}
$\varphi_1$ and $\varphi_2$ transform linearly in the same way,
\begin{eqnarray}
\varphi_1(x)\to \varphi'_1(x')=\Lambda\varphi_1(x) ,
{\bar \varphi_2}(x)\to {\bar \varphi'_2}(x')
={\bar \varphi_2}\beta^0\Lambda^\dagger \beta^0 ,
\end{eqnarray} 
and thus the Lagrangian transforms as
\begin{eqnarray}
L_\varphi\to L'_\varphi=
{\bar \varphi_2}(x)\beta^0\Lambda^\dagger \beta^0 
[i\beta^\mu a_{\mu\nu}D^\nu-m]\Lambda\varphi_1(x). 
\end{eqnarray}
Owing to the invariance of mass term 
$m{\bar\varphi_2}(x)\varphi_1(x)$,  
$\beta^0\Lambda^\dagger \varphi^0=\Lambda^{-1}$. Therefore,
\begin{eqnarray}
T(x)\equiv L'_\varphi-L_\varphi={\bar \varphi_2}(x)i
[\Lambda^{-1}\beta^\mu a_{\mu\nu}D^\nu-\beta^\mu D_\mu]\varphi_1(x).
\end{eqnarray}
Obviously, the conclusion of Lorentz invariance of the action $\int
L_\varphi d^x$ can be drawn once
\begin{eqnarray}\label{t0}
\int T(x)d^4x=\int L'_\varphi d^4x'-\int L_\varphi d^4x=0.
\end{eqnarray}
Under the infinitesimal Lorentz transformation, we have
\begin{eqnarray}
a_{\mu\nu}=g_{\mu\nu}+\varepsilon_{\mu\nu}, &
\Lambda=1-{i\over 2}\varepsilon^{\mu\nu}s_{\mu\nu},
\end{eqnarray}
where $g_{\mu\nu}$ is the metric tensor, $\varepsilon_{\mu\nu}$ is the
infinitesimal antisymmetric
tensor and $s_{\mu\nu}$ is
the unknown coefficient. With the help of $\varepsilon_{\mu\nu}\beta^\mu
D^\nu={1\over 2}\varepsilon_{\rho\tau}(\beta_\rho D_\tau-\beta_\tau
D_\rho)$, $T(x)$ becomes 
\begin{eqnarray}\label{tx}
T(x)={i\over 2}\varepsilon^{\rho\tau}{\bar \varphi_2}(x)[(\beta_\rho
D_\tau-\beta_\tau
D_\rho)-i[\beta^\sigma,s_{\rho\tau}]D_\sigma]\varphi_1(x).
\end{eqnarray}
To prove Eq. (\ref{t0}), all possible cases of Eq. (\ref{tx}) are 
discussed as follows:
\par (1) As $\rho=\tau$, $\varepsilon^{\rho\tau}=0$ and hence
$T(x)=0$; 
\par (2) As $\rho=l$ and $\tau=m$, namely, in the spatial rotation case,
we have  $s_{lm}=-s_{ml}=\epsilon_{lmn}\left ( \begin{array}{cc} \tau_n  0
\\ 0 \tau_n \end{array}\right )$, where $\tau_n$ has been given by Eq.
(\ref{ma}) and $\epsilon_{lmn}$ is the full antisymmetric
tensor($\epsilon_{123}=1$), so $T(x)=0$. As we can see,
$S^2_{23}+S^2_{31}+S^2_{12}=2$, the Dirac-like field thus has spin 1.
\par (3) As $\rho=l$ and $\tau=0$, that is, in the general Lorentz boost
case, the situation is a little more complicated compared with that in  
previous cases. In the present case, $s_{l0}=-s_{0l}={i\over
2}\beta^0\beta^l\equiv{i\over 2}\alpha^l$, and $T(x)$ becomes
\begin{eqnarray}\label{tx}
\begin{array}{ll}
T(x)&={i\over 2}\varepsilon^{l0}\varphi^\dagger _2(x)[-D_l+{1\over
2}(\alpha^m\alpha^l+\alpha^l\alpha^m)D_m]\varphi_1(x) \\
 &=-{i\over 4}\varepsilon^{l0}[\nabla\cdot (\vec{F'}^\dagger F_l)
+\nabla\cdot (\vec{G'}^\dagger G_l)] ,
\end{array}
\end{eqnarray}
where $\vec{F}'=(F'_1, F'_2, F'_3)$(it is similar for $\vec{G}'$),
$\vec{F}$ and $\vec{G}$, the definition of them are
\begin{eqnarray}
\begin{array}{cc}
\varphi_1(x)=\left (
\begin{array}{lcr}
F(x)\\
iG(x)
\end{array} \right ), &
\varphi_2(x)=\left (
\begin{array}{lcr}
F'(x)\\
iG'(x)
\end{array} \right ),
\end{array}
\end{eqnarray} 
in which $F=\left ( \begin{array}{l} F_1 \\ F_2 \\ F_3
\end{array} \right )$, $G$, $F'$ and $G'$ are in a similar form. Since 
$\varphi_1(x)$ and $\varphi_2(x)$ contain only the transversal parts,
they satisfy transversal conditions 
\begin{eqnarray}\label{tc1}
\vec{D}\cdot\vec{F}=\vec{D}\cdot\vec{G}=\vec{D}\cdot\vec{F'}= 
\vec{D}\cdot\vec{G'}=0 
\end{eqnarray}
or 
\begin{eqnarray}\label{tc2}  
\vec{D}^\ast\cdot\vec{F}^\dagger=\vec{D}^\ast\cdot\vec{G}^\dagger
=\vec{D}^\ast\cdot\vec{F'}^\dagger=\vec{D}^\ast\cdot\vec{G'}^\dagger=0.
\end{eqnarray}
In fact, by using Eq. (\ref{f1}) and Eq. (\ref{et}), we can obtain Eq.
(\ref{tc1}) and Eq. (\ref{tc2}) with $A_\mu=0$. 
\par It can be verified also that
\begin{eqnarray}\label{2d}
(\alpha^m\alpha^l+\alpha^l\alpha^m)D_m=(2D_l-M_l-M^T_l),
\end{eqnarray}
where $M^T_L$ is the transpose of $M_L$ and
\begin{eqnarray}
\begin{array}{ccc}
M_1=I_{2\times 2}\bigotimes\left (
\begin{array}{lcr}
D_1 & 0 & 0 \\  
D_2 & 0 & 0 \\
D_3 & 0 & 0   
\end{array} \right ), &
M_2=I_{2\times 2}\bigotimes\left (
\begin{array}{lcr}   
0 & D_1 & 0 \\
0 & D_2 & 0 \\  
0 & D_3 & 0
\end{array} \right ), &
M_3=I_{2\times 2}\bigotimes\left (
\begin{array}{lcr}
0 & 0 & D_1 \\
0 & 0 & D_2 \\
0 & 0 & D_3
\end{array} \right ).
\end{array} 
\end{eqnarray}
Taking use of Eq. (\ref{tc1}), (\ref{tc2}) and (\ref{2d}), we obtain Eq.
(\ref{tx}). As we know, the physical
field vanishes as $|\vec{x}|\to \infty$, so
\begin{eqnarray}
\int T(x)d^4x=-{i\over 4}\varepsilon^{l0}\int
[\nabla\cdot(\vec{F'}^\dagger
F_L)+\nabla\cdot(\vec{G'}^\dagger G_l)]d^4x=0.
\end{eqnarray}
After all the possible cases (1), (2) and (3) have been considered
, the conclusion $\int T(x)d^4x\equiv 0$ becomes obvious. That is
to say, the action of transversal Dirac-like field is Lorentz invariant.
\par Since the action $\int L_\varphi d^4x$ is Lorentz invariant while the
corresponding Lagrangian $L_\varphi$ is not, the Lorentz invariance has
a special implication in our theory. We will discuss it further in
the next section. 
\section{Feynman rules and Polarization cross section for $e^+ e^-\to f^+
f^-$}
\indent
\par Under the gauge transformation $\varphi_1(x)\to
e^{i\theta(x)}\varphi_1(x)$(where
$\theta(x)$ is a real parameter), the field quantity $\varphi_2(x)$
transforms in the same way for the reason that the positive and the
negative frequency parts of $\varphi_1(x)$ or $\varphi_2(x)$ are linearly
independent. Obviously, the Lagrangian given by Eq. (\ref{lag1}) is
invariant under the global gauge transformation and the corresponding
conserved charge is
\begin{eqnarray}
Q=\int \varphi^\dagger_2(x)\varphi_1(x)d^3x=\sum\limits_{\vec{p},s}
[a^\dagger(\vec{p},s)a(\vec{p},s)-b^\dagger(\vec{p},s)b(\vec{p},s)],
\end{eqnarray}
just as what we expect. Minimal electromagnetic coupling is easily
introduced into Eq. (\ref{lag1}) by making the replacement
$\partial_\mu\to
D_\mu=\partial_\mu-ieA_\mu$ and Eq. (\ref{lag1}) turns into Eq.
(\ref{lag2}). One can verify that Eq. (\ref{lag2}) is invariant
also under local gauge transformation
\begin{eqnarray}
\left\{\begin{array}{ll}
&\varphi_1(x)\to e^{i\theta(x)}\varphi_1(x) 
(hence~\varphi_2(x)\to e^{i\theta(x)}\varphi_2(x)) \\
&A_\mu (x)\to A_\mu (x)-{1\over e}\partial_\mu\theta(x).
\end{array}
\right.
\end{eqnarray}
\par Now, we develop the relevant perturbative
theory(called "vector QED" or "VQED") in adiabatic approximation, where
the S-matrix can be calculated from Dyson's formula
\begin{eqnarray}\label{act}
S=\sum\limits^{\infty}_{n=0}S^{(n)}=\sum\limits^{\infty}_{n=0}
{(-i)^n \over n!}\int d^4x_1\cdots d^4x_n T\{H_I(x_1)\cdots H_I(x_n)\},
\end{eqnarray}  
where $H_I(x)$ is the interaction Hamiltonian written in interaction
picture. After a tedious calculation, we give the Feynman rules for VQED
in momentum space as follows(for simplicity, the Dirac-like particle or
antiparticle is denoted by DL or anti-DL, respectively):
\par 1. Propagators:
\par Photon's$=\langle T A_\mu(x_1)A_\nu(x_2)\rangle
={-ig_{\mu\nu}\over q^2+i\varepsilon}$,
\par DL's$=\langle T \varphi_1(x_1)\bar \varphi_2(x_2) \rangle=
{iA_\bot \over \beta\cdot p-m+i\varepsilon}$.
\par 2. External lines:
\par Photon annihilation$=\epsilon_\mu(p)$, 
Photon creation$=\epsilon^\ast_\mu(p)$.
\par DL annihilation$=\sqrt{{m\over VE}}\chi(p,s)$, 
DL creation$=\sqrt{{m\over VE}}\bar\chi(p,s)$.
\par anti-DL annihilation$=\sqrt{{m\over VE}}\bar y(p,s)$,
anti-DL creation$=\sqrt{{m\over VE}}y(p,s)$. 
\par 3. Vertex(DL with photon): $-ie\beta_\mu$.
\par 4. Impose momentum conservation at each vertex.
\par 5. Integrate over each free loop momentum: $\int {d^4p\over 
(2\pi)^4}$.
\par 6. Divide by symmetry factor.
\par As an application, we will calculate the polarization cross section
for process $e^-(p,s)+e^+ (q,t)\to f^-(p',s')+f^+ (q',t')$: the
annihilation of an electron with a positron to create a pair of
Dirac-like particles
$f^+$ and $f^-$. For simplicity, we work in the center-of-mass(CM) frame
and compute the relevant transition amplitude in the lowest order.
\par In our case, the initial state is $|e\rangle=c^\dagger
(p,s)d^\dagger (q,t)|0\rangle$ and the final state is $|f\rangle=a^\dagger
(p',s')b^\dagger (q',t')|0\rangle$, where $c^\dagger $, $d^\dagger $
are the creation operators of electron and positron and $s$, $t(=1,2)$ are
their spin indices.
$s'$, $t'(=\pm 1)$ are the transversal polarization indices while $p$,
$q$, $p'$ and $q'$ are the 4-momentum. The interaction Hamiltonian related
to the process is
\begin{eqnarray}
H_I(x)=e{\bar\psi}(x)\gamma^\mu\psi(x)A_\mu(x)+e\bar
\varphi_2(x)\beta^\mu\varphi_1(x)A_\mu(x),
\end{eqnarray}
where $\gamma^\mu$ is Dirac matrix and $\bar
\psi(x)=\psi^\dagger(x)\gamma^0$, in which
\begin{eqnarray}
\psi(x)=\sum\limits_{\vec{p},s}\sqrt{{m_0\over VE_0}}
[c(p,s)u(p,s)e^{-ipx}+d^\dagger(p,s)\nu(p,s)e^{ipx}]
\end{eqnarray}
is the electronic field, $m_0, E_0$ are the mass and the energy of
electron, respectively. Then the corresponding transition amplitude is
\begin{eqnarray}
\begin{array}{lll}
S^{(2)}_{fe}&=&\langle f|S^{(2)}|e\rangle\\
&=&(2\pi)^4\delta^4(p+q-p'-q')\sqrt{{m\over vE}}{\bar
\chi}(p',s')(-ie\beta_\mu)\sqrt{{m \over vE}}y(q',t')\\
&&{-ig^{\mu\nu}\over (p+q)^2} 
\sqrt{{m_0\over vE_0}}{\bar \nu}(q,t)(-ie\gamma_\nu)
\sqrt{{m_0\over vE_0}}u(p,s),
\end{array}
\end{eqnarray}
where $S^{(2)}$ is defined by Eq. (\ref{act}).
\par However, since $e^+$ and $e^-$ have spin 1/2 while $f^+$ and $f^-$
have spin 1, conservation of angular momentum requires that both the total
spin of $e^+e^-$ and that of $f^+f^-$ are zero, which implies that
both the initial state current(denoted by $J^\mu_0$) and final state
current(denoted by $J^\mu$) must be spin singlets, so the relevant cross
section is a polarization one(denoted by $\sigma_{polar}$). In the CM
frame, 
\begin{eqnarray}
J^\mu_0={1\over \sqrt{2}}[{\bar \nu}(q,1)\gamma^\mu u(p,1)
-{\bar \nu}(q,2)\gamma^\mu u(p,2)],\\
J^\mu={1\over \sqrt{2}}[{\bar \chi}(p',1)\beta^\mu y(q',1)
-{\bar \nu}(p',-1)\beta^\mu y(p,-1)].
\end{eqnarray}
It is not difficult to infer that
\begin{eqnarray}
\sigma_{polar}=&\int (2\pi)^4\delta^4(p+q-p'-q'){m^2_0m^2\over
2|\vec{p}|E_0E^2}
{e^4\over (p+q)^4}{d^3p'\over (2\pi)^3}{d^3q'\over (2\pi)^3}M^2,
\end{eqnarray}
where 
\begin{eqnarray}
M^2=|J_{\mu}J^{\mu}_0|^2.
\end{eqnarray}
Obviously, ${M\over (p+q)^2}$ corresponds to the invariant transition
amplitude. As a result, $M^2$ must be Lorentz invariant. After a
cumbersome calculation, we obtain
\begin{eqnarray}
M^2=4\cos^2\theta,
\end{eqnarray}
where $\theta$ stands for the angle between the incoming electrons and the
outgoing Dirac-like particles. Lorentz invariance of $M^2$ requires that
$\theta=0$ or $\pi$(as we know, only in these cases, the $\cos^2\theta$ is
Lorentz invariant). Namely, the final state momentum is purely
longitudinal. Therefore, the Lorentz invariance has an incidental meaning
for our theory(i.e. VQED): it provides an additional constraint for a
possible physical process, just as the conservation of quantum number
provides a selection rule for a possible particle reaction. In other
words, VQED contains not only the Lorentz invariant process(and hence
this process is allowed), but also the process violating the Lorentz
invariance and hence being forbidden. In a sense, this paper provides a
new approach to construct certain quantum field theories still unknown to
us.   
\par In a word, $M^2=4$. Seeing that $E_0=E$ in the CM frame and
$|\vec{p}|\approx E_0$ in the high energy approximation, we obtain
\begin{eqnarray}
\sigma_{polar}={\pi\alpha^2m^2_0m^2\over E^6_0}v'^3,
\end{eqnarray}
where $\alpha={e^2 \over 4\pi}$ and $v'$ is the velocity of $f^+$(or
$f^-$). It is found that $v'={\sqrt{3}\over 3}$ maximizes
$\sigma_{polar}$, that is 
\begin{eqnarray}
(\sigma_{polar})_{max}={2\sqrt{3}\pi\alpha^2\over 243}{m^2_0\over m^4}.
\end{eqnarray} 
While in the high energy limit, $v'\approx 1$, and
\begin{eqnarray}
\sigma_{polar}={\pi\alpha^2m^2_0m^2\over E^6_0}.
\end{eqnarray}
\par To sum up, the following conclusions follows:
\par (1), In the high energy approximation, $m, m_e\ll E_0=E$,
$\sigma_{polar}\approx 0$.
\par (2), The momentum of $f^+$ and $f^-$ is purely longitudinal.
\par In the light of (1) and (2), we can explain the fact that none of the
Dirac-like particles have been found so far.
\section{Renormalizability of VQED}
\indent
\par After the introduction of VQED in Sec. VII, the question whether
VQED is renormalizable then arises. Before proceeding discussions, let
us pause a moment to mention the following fact: Let the Dirac-like field
quantity $\varphi=\left(\begin{array}{l} F\\ iG \end{array} \right )$ and
set $m=0$, where $F=\left(\begin{array}{l}
F_1\\F_2\\F_3\end{array}\right)$ and
$G=\left(\begin{array}{l}G_1\\G_2\\G_3\end{array}\right)$. In term of F
and G, Eq. (\ref{dl}) or Eq. (\ref{dl2}) can be reexpressed as Maxwell
equations written in free vacuum, where F and G correspond to the electric
field intensity $E'$ and the magnetic field intensity $H'$, respectively.
However, we can't regard F and G directly as E' and H' correspondingly,
for two reasons:
\par (1), Under an infinitesimal Lorentz boost parametrized by the
infinitesimal velocity v, the transformation properties of F(or G) are
different from those of E'(or H'). The transformation of F(or G) contains
the factor of ${v\over 2}$ while the transformation of E'(or H') contains
the factor v.
\par (2), The dimension of F(or G) is ${3\over 2}$ while that of E'(or H')
is 2. In fact, the canonical dimension of the Dirac-like vector is
different from that of electromagnetic field.
\par In a word, the Dirac-like field is a new kind of vector field.
\par Now, let us pay attention to the renormalizability mentioned above.
Fortunately, VQED can be developed in term of the same formalism as
those in the electron-photon interaction theory, namely, spinor
QED(denoted by SQED).
In particular, the free propagator and the vertex in VQED take the same
forms as those in SQED except for involving matrices $\beta^\mu$(or
$\beta^\mu A_\bot$) instead of the Dirac matrices $\gamma^\mu$. As a
result, starting from the Feynman integral(i. e. the loop momentum
integral) of SQED, we can obtain the corresponding Feynman integral of
VQED by replacing the $\gamma^\mu$ with $\beta^\mu$(or $\beta^\mu
A_\bot$). Owing to these facts, the discussions of the
renormalization of VQED is just a step by step business.
\par Let us consider a n-vertex one-particle-irreducible(1PI) diagram in
VQED, in which there are N external Dirac-like particle lines and N'
external photon lines. Taking use of the similarity between VQED and
SQED\cite{pest}, we obtain the superficial degree of divergence of the 1PI
diagram, say D, as follow:
\begin{eqnarray}
D=4-N'-{3\over 2}N,
\end{eqnarray} 
from which we draw the following conclusions:
\par 1, D is independent of the number n of the vertices(i.e. the order n
of perturbation).
\par 2, D depends only on the number of external lines of the 1PI diagram,
and there are only a finite number of external lines with $D\geq 0$.
\par From (1) and (2), we come to the conclusion that VQED is
renormalizable.

\end{document}